\documentclass[conference]{IEEEtran}
\usepackage{fancyhdr}
\IEEEoverridecommandlockouts

\newif\ifarxiv
\arxivtrue

\usepackage{cite}
\usepackage{amsmath,amssymb,amsfonts}
\usepackage{graphicx}
\usepackage{textcomp}
\usepackage{xcolor}
\usepackage{booktabs}
\usepackage{tikz}
\usetikzlibrary{arrows.meta,positioning,fit}
\usepackage{listings}
\usepackage{url}

\graphicspath{{figures/}}

\begin{document}

\title{No-Restart Elasticity in an Adaptive Runtime System for Cloud-Native HPC}

\author{
\IEEEauthorblockN{Aditya Bhosale}
\IEEEauthorblockA{\textit{University of Illinois Urbana-Champaign}\\
Urbana, IL\\
adityapb@illinois.edu}
\and
\IEEEauthorblockN{Laxmikant Kale}
\IEEEauthorblockA{\textit{University of Illinois Urbana-Champaign}\\
Urbana, IL\\
kale@illinois.edu}
}

\maketitle
\thispagestyle{fancy}
\lhead{}
\rhead{}
\chead{}
\ifarxiv
\lfoot{\footnotesize{\parbox{\textwidth}{\copyright\ 2026 IEEE. Personal use of this material is permitted. Permission from IEEE must be obtained for all other uses, in any current or future media, including reprinting/republishing this material for advertising or promotional purposes, creating new collective works, for resale or redistribution to servers or lists, or reuse of any copyrighted component of this work in other works.}}}
\else
\lfoot{\footnotesize{
SC26 Workshops, November 15-20, 2026, Chicago, Illinois, USA
\newline 979-8-3195-1221-5/26/\$31.00 \copyright 2026 IEEE}}
\fi
\rfoot{}
\cfoot{}
\renewcommand{\headrulewidth}{0pt}
\renewcommand{\footrulewidth}{0pt}

\begin{abstract}
Exploiting discounted spot instances for HPC requires an application to change its resource allocation at runtime, shrinking ahead of an interruption and expanding onto replacement capacity. Existing elasticity mechanisms implement rescaling as a full process teardown followed by a cold restart at the new processor count, and the restart stage accounts for up to 95\% of the total overhead, growing with node count and increasing substantially on GPUs due to CUDA context initialization. In this paper, we present a no-restart rescaling mechanism for the Charm++ runtime system in which surviving processes never exit: a rescaling operation consists of a membership negotiation with a lightweight external coordinator, reconciliation of the UCX communication endpoints with the new cluster view, and a return to the top of the runtime initialization path that preserves live application state. This reduces the cost of a rescaling operation, excluding the load balancing step that any rescaling model requires, from multiple seconds to 8--15\,ms on CPUs and 7--11\,ms on GPUs, at 4 to 32 instances. Because processes survive, GPU device state persists in place, eliminating the checkpointing daemons previously required for GPU elasticity, and a launcher-independent bootstrap mechanism removes the dependence on supervised process managers, which are incompatible with spot instance interruptions. Integrated with an existing spot instance management framework, the mechanism cuts the end-to-end overhead of eight simultaneous interruptions to 0.2\% of runtime on CPUs and 0.6\% on GPUs, and raises the rate at which a job can be rescaled below 1\% overhead by a factor of seven.
\end{abstract}


\section{Introduction}
\label{sec:introduction}

The convergence of HPC and cloud computing has accelerated in recent years. Cloud providers now offer HPC-oriented instance types and high-performance interconnects, while the capital cost of dedicated supercomputers and the growth of AI/ML workloads continue to drive HPC users toward cloud platforms. Among the strongest economic arguments for the cloud is the spot market: providers sell spare capacity at discounts of 60--90\% under the condition that instances may be interrupted with only a short warning when demand for on-demand capacity surges. Prior work has shown that adaptive runtime systems such as Charm++ can exploit spot instances for tightly coupled HPC applications by rescaling the application in response to interruptions, shrinking it onto the surviving instances and expanding it again when replacement capacity arrives, achieving substantial cost savings~\cite{bhosale2025hpdc}, and that proactive replacement policies can hide most of the capacity loss associated with an interruption~\cite{bhosale2026ashes}.

A fundamental inefficiency underlies this prior work: every rescaling operation is implemented as a restart. Whether the application state is checkpointed to a shared filesystem or to Linux shared memory, changing the number of processing elements (PEs) requires terminating every process in the job and launching a new set of processes at the new PE count, which then restore the saved state. Results from previous work~\cite{bhosale2026ashes} show that this restart stage dominates the total cost of a rescaling operation. Of an approximately 2.3\,s shrink operation on CPUs, roughly 2\,s is spent in process restart; on GPUs, the restart cost rises to 6--7\,s of approximately 7.5\,s total because every new process must reinitialize its CUDA context. The restart cost also grows with the number of nodes, since application launch time scales with job size. In contrast, checkpoint, load balancing, and restore---the stages of a rescaling operation that perform essential work---constitute only a small fraction of the total overhead.

The restart-based model imposes costs beyond wall-clock overhead. On GPUs, process termination destroys the CUDA context, so GPU-resident data must be moved out of the terminating process. Prior work required a per-GPU daemon process and CUDA Interprocess Communication (IPC) machinery solely to preserve device data across the restart~\cite{bhosale2026ashes,bhosale2025sci}. The multi-second overhead also constrains resource management policy: rescaling must be treated as a rare event to be batched and deferred rather than an operation that can be invoked freely.

Moreover, the restart model is the only model that conventional launchers permit. MPI-style launchers such as \texttt{mpirun} and PRRTE maintain a supervised daemon on every host, abort the entire job when any daemon disconnects, and provide no mechanism for adding or removing hosts from a running job. A full teardown followed by a new launch is the only membership change they support, and prior systems relied on exactly this relaunch to rescale. Eliminating the restart therefore requires eliminating the dependence on a supervised launcher as well. The same requirement arises in cloud-native resource management generally, where an orchestrator such as Kubernetes or a cloud fleet manager adds and removes nodes independently of the job and expects the workload to absorb the change; a launcher that treats every departure as fatal cannot be driven this way.

In this paper, we eliminate the restart. We present a rescaling mechanism for Charm++ over UCX in which surviving processes never exit. A shrink or expand operation proceeds as follows: (1)~PE~0 negotiates the new cluster membership with a lightweight external TCP coordinator; (2)~departing processes exit cleanly; (3)~every surviving process reconciles its UCX endpoints with the new membership view and returns control to a point at the top of the runtime entry function, re-executing initialization in a mode that preserves live state; and (4)~joining processes start as new processes but bypass \texttt{main()}, adopting runtime state from a broadcast. The application then resumes its iteration loop. Excluding the load balancing step that evacuates departing PEs before a shrink or populates new PEs after an expansion, which any rescaling model requires, this sequence completes in approximately 10\,ms on our test configuration, a reduction of more than two orders of magnitude relative to the restart-based approach. Because the mechanism is driven entirely by an external request, and a new node joins given only the coordinator's address, the running job presents to an external orchestrator as a pool of nodes that can be grown or shrunk at any time, the interface that cloud-native resource managers expect.

Realizing this model requires addressing a substantial systems problem: every piece of runtime state that was designed to be initialized exactly once per process lifetime must now survive an arbitrary number of reinitializations within a single process lifetime. Handler tables, reduction trees, location managers, load balancer synchronization, and timers all carry state that is either initialization-time state, which must be preserved rather than re-created, or topology-dependent state, which must be adjusted rather than preserved. We describe the lifecycle pattern we developed to manage this distinction and the mechanisms it requires.

This paper makes the following contributions:
\begin{itemize}
\item We present a no-restart shrink/expand mechanism for a message-driven parallel runtime system, based on coordinated membership change, in-place UCX endpoint reconciliation, and reinitialization of surviving processes without process termination. The mechanism reduces rescaling overhead, excluding load balancing, from seconds to approximately 10\,ms (Sections~\ref{sec:design} and~\ref{sec:state}).
\item We present a launcher-independent bootstrap mechanism built on a standalone membership coordinator and an SSH-based launcher that spawns the ranks and exits, leaving no supervising daemons. This removes the process manager dependence that makes conventional launchers incompatible with spot instance interruptions (Section~\ref{sec:design}).
\item We show that the no-restart model provides GPU elasticity without checkpointing daemons: because processes survive rescaling operations, CUDA contexts and device-resident data persist in place, making the CUDA IPC daemon subsystem required by prior work unnecessary (Section~\ref{sec:state}).
\item We integrate the mechanism with an existing spot instance management framework with proactive capacity rebalancing, analyze how low-cost rescaling changes the space of viable interruption-handling policies, and evaluate the system end-to-end on CPU and GPU spot fleets (Sections~\ref{sec:spot} and~\ref{sec:evaluation}).
\end{itemize}

\section{Background and Motivation}
\label{sec:background}

\subsection{Charm++ and Migratable Objects}

Charm++~\cite{kale1993charm,acun2014charm} is an asynchronous, message-driven parallel programming model in which computation is expressed in terms of migratable C++ objects called \emph{chares}, mapped to processing elements (PEs) by the runtime system and communicating through non-blocking entry method invocations delivered by a distributed location management system. Chares are organized into indexed collections; two special collection types, \emph{groups} and \emph{nodegroups}, have exactly one member per PE and per process, and the runtime system itself is largely built from groups, which implement per-PE services such as load balancing, location management, and reductions. Applications also declare \emph{read-only} variables, global data initialized on PE~0 at startup and broadcast to all PEs. Because applications are overdecomposed into many more chares than PEs and the runtime manages object placement, it can migrate chares during execution, enabling measurement-based dynamic load balancing~\cite{acun2014charm} and, most relevant to this work, elasticity (termed malleability in the job scheduling literature): changing the number of PEs available to a running application~\cite{gupta2014malleable}.

\subsection{The Kill-and-Restart Elasticity Model}
\label{sec:killrestart}

Prior implementations of shrink/expand in Charm++ follow a model that we refer to as kill-and-restart~\cite{gupta2014malleable,bhosale2025hpdc}. To shrink, the runtime system first migrates chares away from the PEs to be removed using a load balancing step that also rebalances the remaining load. The application state, consisting of read-only global data, group (per-PE) objects, and all chare data, is then checkpointed into Linux shared memory, itself an optimization over the earlier practice of checkpointing to a shared filesystem, on the surviving hosts. Every process in the job exits, the launcher restarts the application with the new PE count, and the new processes restore their state from the shared-memory checkpoint and resume execution. Expansion is almost symmetric: the runtime checkpoints, restarts at the larger PE count, restores, and performs a load balancing step to populate the new PEs.

For GPU applications, this model requires additional machinery. Process termination destroys the CUDA context and, with it, all device allocations. To avoid staging device data through host memory on every rescaling operation, prior work launched a persistent daemon process for each GPU. At checkpoint time, each PE passed its device buffer pointers to the daemon using CUDA IPC, and the daemon copied the data into buffers that it owned. After restart, the daemon passed the data back to the application processes~\cite{bhosale2026ashes,bhosale2025sci}. This approach kept checkpoint and restore traffic on the device but introduced a daemon lifecycle that must be managed on every host, including replacement instances.

\subsection{Spot Instances and CharmCloudManager}
\label{sec:ccm-background}

Cloud providers sell spare capacity as spot instances at significant discounts, with the caveat that instances can be interrupted after a short warning (two minutes on AWS EC2). \texttt{CharmCloudManager}~\cite{bhosale2025hpdc} manages Charm++ applications on mixed on-demand and spot fleets: a monitoring task watches for interruption notices and newly launched instances, maintains the nodelist, and signals the application to shrink before an interruption takes effect and to expand when replacement capacity arrives. The fleet must include at least one on-demand instance to host PE~0, which coordinates every rescaling operation and therefore must never be interrupted. Prior work~\cite{bhosale2026ashes} added capacity rebalancing, launching a replacement proactively when an instance is flagged at elevated risk so that the application rescales once, directly onto the replacement, rather than shrinking on interruption and expanding later; this halved the interruption-handling overhead by eliminating one of the two restarts.

\subsection{Motivation: Restart Dominates Rescaling Cost}
\label{sec:motivation}

\begin{table}[t]
\centering
\caption{Breakdown of shrink overhead under the kill-and-restart model when one instance is interrupted, from~\cite{bhosale2026ashes} (Jacobi2D, $16{,}384 \times 16{,}384$ grid, 16 instances).}
\label{tab:restart-dominance}
\begin{tabular}{lrr}
\toprule
Stage & CPU (c6gn.large) & GPU (g4dn.xlarge) \\
\midrule
Checkpoint    & $\sim$0.05\,s & $\sim$0.05\,s \\
Load balance  & $\sim$0.15\,s & $\sim$0.35\,s \\
Restart       & $\sim$1.9\,s  & $\sim$6.8\,s  \\
Restore       & $\sim$0.05\,s & $\sim$0.05\,s \\
\midrule
Total         & $\sim$2.2\,s  & $\sim$7.2\,s  \\
\bottomrule
\end{tabular}
\end{table}

Table~\ref{tab:restart-dominance} summarizes the stage-by-stage overhead of a shrink operation under the kill-and-restart model, as measured in prior work~\cite{bhosale2026ashes}. Two observations motivate the present work.

First, restart is the dominant cost. In-memory checkpointing already reduced the checkpoint and restore stages to tens of milliseconds, and the load balancing cost is modest and unavoidable. Process teardown and cold restart account for 85--95\% of the total overhead. Moreover, unlike the other stages, the restart cost increases with job size, because application launch time grows with the number of nodes. On GPUs, the restart penalty approximately triples due to CUDA context initialization, which requires multiple seconds per process. Any further significant reduction in rescaling overhead must therefore come from the restart stage.

Second, the restart model is imposed by the launchers it runs on. \texttt{mpirun} and PRRTE maintain a supervised daemon on every host, treat the loss of any daemon as a fatal job event, and provide no mechanism for changing the host set of a running job. A full teardown followed by a relaunch is therefore the only membership change they support. Prior work rescaled through exactly this relaunch, which incidentally protected the job from launcher failures: every process, and the launcher session itself, terminated before any host disappeared. This protection is unavailable to a no-restart design, in which surviving processes must continue executing while interrupted hosts disappear, a condition that supervised launchers treat as fatal. Launcher independence must therefore be addressed as an explicit design requirement.
\section{Design}
\label{sec:design}

\subsection{Overview}
\label{sec:design-overview}

The central idea of our design is that a rescaling operation should be a reconfiguration of live processes rather than a restart of the job: PE~0 negotiates the new membership with an external TCP coordinator; departing PEs evacuate their chares through a load balancing step and exit; survivors reconcile their UCX endpoints with the new view and \texttt{longjmp} back to the runtime entry function, re-executing initialization in a mode that skips or adapts work that must not repeat (Section~\ref{sec:state}); newcomers bypass \texttt{main()} and adopt a broadcast of read-only and group state from PE~0; and a recorded resume callback re-enters the application. No process image is reloaded, no CUDA context is destroyed, and no state is written to a filesystem or shared memory.

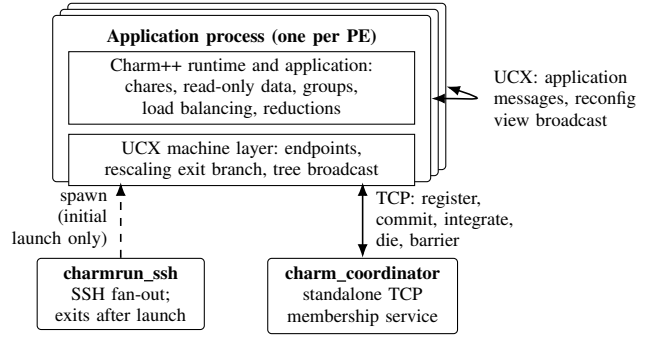
\begin{figure}
\centering
\begin{tikzpicture}[
  font=\scriptsize,
  comp/.style={draw, rounded corners=1.5pt, align=center, inner sep=3pt,
               fill=white},
  layer/.style={draw, align=center, inner sep=2.5pt, fill=white,
                text width=44mm},
  arr/.style={-{Latex[length=1.8mm]}, semithick},
  dblarr/.style={{Latex[length=1.8mm]}-{Latex[length=1.8mm]}, semithick},
]

\node[comp, minimum width=50mm, minimum height=22mm]
  (procBack2) at (2.0mm, 1.6mm) {};
\node[comp, minimum width=50mm, minimum height=22mm]
  (procBack1) at (1.0mm, 0.8mm) {};
\node[comp, minimum width=50mm, minimum height=22mm] (proc) at (0,0) {};

\node[anchor=north, font=\scriptsize\bfseries]
  at ([yshift=-0.6mm]proc.north)
  {Application process (one per PE)};

\node[layer, anchor=north] at ([yshift=-4.8mm]proc.north) (appLayer)
  {Charm++ runtime and application:\\
   chares, read-only data, groups,\\
   load balancing, reductions};

\node[layer, anchor=north] at ([yshift=-1.2mm]appLayer.south) (ucxLayer)
  {UCX machine layer: endpoints,\\
   rescaling exit branch, tree broadcast};

\draw[dblarr] (proc.east) .. controls +(6mm,0) and +(6mm,-2mm) ..
  (procBack2.east)
  node[midway, right=1.5mm, align=left]
  {UCX: application\\ messages, reconfig\\ view broadcast};

\node[comp, text width=20mm, anchor=north]
  at ([xshift=-16mm,yshift=-10mm]proc.south) (launcher)
  {\textbf{charmrun\_ssh}\\
   SSH fan-out;\\
   exits after launch};

\node[comp, text width=23mm, anchor=north]
  at ([xshift=16mm,yshift=-10mm]proc.south) (coord)
  {\textbf{charm\_coordinator}\\
   standalone TCP\\
   membership service};

\draw[arr, dashed] (launcher.north) --
  ([xshift=-16mm]proc.south)
  node[midway, left=0.5mm, align=right] {spawn\\ (initial\\ launch only)};

\draw[dblarr] (coord.north) --
  ([xshift=16mm]proc.south)
  node[midway, right=0.5mm, align=left]
  {TCP: register,\\ commit, integrate,\\ die, barrier};

\end{tikzpicture}
\caption{System components. The coordinator is the only persistent external process. The launcher spawns the initial cluster and plays no further role in execution. Survivors learn reconfiguration views over UCX; only PE~0, removed members, and newcomers communicate with the coordinator during a rescaling operation.}
\label{fig:components}
\end{figure}

Three kinds of processes participate in the system (Figure~\ref{fig:components}): the application binary, with one process per PE; a single standalone coordinator process that maintains the canonical cluster membership; and a launcher that spawns the initial cluster and subsequently plays no role. The coordinator is deliberately built without any Charm++ dependencies; it is a small TCP server that is independent of the runtime system.

\subsection{The Coordinator}
\label{sec:coordinator}

The coordinator maintains the canonical cluster \textit{view}, consisting of an epoch number and an ordered member list in which each entry carries a node ID and the corresponding node's UCX worker address. The epoch number is a monotonically increasing counter that identifies a version of the membership; it is incremented on every committed reconfiguration. Protocol messages carry the epoch of the view they refer to, which allows the coordinator and the ranks to detect requests and barrier arrivals that belong to an outdated view. The coordinator implements a small message protocol over TCP, organized into four message flows.

\subsubsection{Initial registration} Each rank of the initial launch sends a \textsc{Register\_Initial} message containing its launcher-assigned node ID and its UCX address. Once the expected number of ranks has registered, the coordinator replies to all of them with the complete member list. This exchange provides the address distribution that a process management interface (PMI)~\cite{balaji2010pmi,castain2018pmix} would otherwise perform.

\subsubsection{Newcomer registration} A process joining during an expansion sends a \textsc{Register\_Newcomer} message with its UCX address and receives an immediate snapshot of the current cluster. The process is then queued in FIFO order until a commit consumes it, at which point the coordinator sends it an \textsc{Integrate} message containing its assigned node ID and the final post-rescale view.

\subsubsection{Commit} PE~0 drives every reconfiguration. It first issues a \textsc{Query\_Pending} message to determine how many newcomers are queued, and then sends a \textsc{Commit} message containing the epoch, the set of nodes to remove, and the number of newcomers to absorb. The coordinator compactly renumbers the survivors, appends the consumed newcomers, increments the epoch, sends a \textsc{Die} message to each removed member and an \textsc{Integrate} message to each consumed newcomer, and returns the new view to PE~0 in a \textsc{Commit\_Reply} message.

\subsubsection{Barrier} The coordinator provides a barrier across all live ranks, used during job shutdown under coordinator bootstrap (Section~\ref{sec:bootstrap}), where it substitutes for the barrier that PMI would otherwise provide at termination.

An important scalability property of the design is that survivors do not learn the new view over TCP. A straightforward alternative, in which the coordinator pushes the view to every survivor, serializes an $O(N)$ fan-out through a single TCP server. Instead, PE~0 serializes the view into a reconfiguration message and forwards it over UCX along a binary tree rooted at itself, each survivor applying the view delta and forwarding to its children. The closing synchronization follows the same principle, a barrier over the new endpoints along the same tree. The coordinator therefore communicates only with PE~0, the removed members, and the consumed newcomers, and its work is independent of the number of survivors.

\subsection{Bootstrap Modes and Launcher Independence}
\label{sec:bootstrap}

The initialization routine of the UCX machine layer supports three entry paths.

\subsubsection{Survivor reinitialization} After the \texttt{longjmp}, a survivor's machine layer initialization performs almost no work: it reports the new node count and the process's new node ID, both of which were recorded before the jump. The endpoints were already reconciled with the new view in the rescaling branch of the exit path, so communication is available as soon as initialization returns.

\subsubsection{Newcomer} A process launched with the \texttt{+newcomer} flag initializes UCX locally, registers with the coordinator, and receives a snapshot of the current cluster. It then speculatively opens endpoints to the current members and blocks awaiting the \textsc{Integrate} message; the endpoint handshakes complete while the application continues to run. When the final view arrives, the newcomer compares its speculative endpoints against the committed membership, creating, retaining, or closing endpoints as needed, and joins the survivors at a tree barrier.

\subsubsection{Initial launch} Under coordinator bootstrap, the launcher passes each rank its node ID, the node count, and the coordinator address on the command line; ranks register with the coordinator and build endpoints directly from the returned member list, without involving a process management interface. For non-elastic runs, conventional launchers remain supported, with PMI providing rank assignment and address exchange.

Coordinator bootstrap is what makes elastic execution possible. Supervised launchers such as \texttt{mpirun} treat any daemon disconnection as fatal: an interrupted spot host, or even a host that is cleanly vacated after a shrink, terminates the job. PRRTE has less aggressive daemon-loss semantics but still aborts on TCP connection close. Retaining PMI without the supervised daemons is not an option, because PMI is an interface to the process manager rather than an independent service: its server side is hosted by those daemons, which supply rank assignment and the key-value store~\cite{balaji2010pmi,castain2018pmix}. Abandoning the supervised daemon model therefore requires a different provider for both roles. The coordinator, which must exist anyway for membership management, assumes them. We therefore developed \texttt{charmrun\_ssh}, a launcher that fans out over SSH, spawns the ranks, and exits. The disappearance of a vacated host has no effect on the surviving cluster; the only component that must remain reachable is the coordinator.

\subsection{The Shrink Flow}
\label{sec:shrink}

A shrink is triggered by an external request. In our system, the resource manager delivers a keep/remove bitmap over the Charm++ Converse Client-Server (CCS) remote request interface to PE~0 (Section~\ref{sec:spot}). The request is folded into the next load balancing step: the load balancer computes a migration plan that evacuates all chares from the departing PEs while rebalancing load across the survivors, and the migrations execute normally. When migration completes, the runtime enters the rescaling path:

\begin{enumerate}
\item The load balancer migrates all chares from the departing PEs while rebalancing load across the survivors. The rescaling path begins once migration completes.
\item A broadcast marks every PE as participating in a rescaling operation. This flag, set on all PEs, is the only condition the runtime checks to enable rescaling-specific behavior (Section~\ref{sec:state}). A barrier then ensures that every PE has entered the rescaling path before the reconfiguration proceeds.
\item Each PE computes its post-rescale node ID from the bitmap and enters the runtime exit path, which detects the rescaling operation and branches away from process termination.
\item PE~0 queries the coordinator for pending newcomers (of which there are none for a pure shrink) and then commits the removal set. Departing PEs receive the \textsc{Die} message, close their coordinator connections, and exit. The remaining survivors receive the new view through the UCX tree broadcast originating at PE~0.
\item Every survivor reconciles its endpoint table with the new view: endpoints between surviving pairs are retained and re-indexed under the new numbering (Section~\ref{sec:tags}), and endpoints to removed members are closed. A pure shrink creates no endpoints. Each survivor then joins a tree barrier executed over the new endpoints; its completion guarantees that every rank has finished reconciliation before any application traffic flows, and it constitutes the first exercise of the new endpoints.
\item Each survivor executes \texttt{longjmp} to the \texttt{setjmp} point in the runtime entry function. Runtime initialization executes again in survivor mode, PE~0 broadcasts read-only and group state, and the recorded resume callback re-enters the application.
\end{enumerate}

\subsection{The Expand Flow}
\label{sec:expand}

\begin{figure}
\centering
\begin{tikzpicture}[
  font=\tiny,
  tcp/.style={-{Latex[length=1.5mm]}, semithick},
  ucx/.style={-{Latex[length=1.7mm]}, very thick},
  lane/.style={draw, fill=white, align=center, font=\tiny\bfseries,
               inner sep=2pt, minimum height=4.5mm, minimum width=13mm},
  act/.style={draw, fill=white, align=center, inner sep=1.5pt},
  x=1mm, y=1mm,
]
\def\xN{7} \def\xC{28} \def\xP{48} \def\xS{66} \def\xD{82}

\fill[gray!12] (\xP-5,-4) rectangle (\xD+4,-21);
\node[anchor=west, gray, font=\tiny\itshape] at (\xP-4.5,-6)
  {application running};
\fill[gray!12] (\xN-5,-84) rectangle (\xS+4,-90);
\node[gray, font=\tiny\itshape] at ({(\xN+\xS)/2},-87)
  {application resumes at new size};

\node[lane] at (\xN,0) {Newcomer};
\node[lane] at (\xC,0) {Coordinator};
\node[lane] at (\xP,0) {PE 0};
\node[lane] at (\xS,0) {Survivors};
\node[lane] at (\xD,0) {Departing PEs};

\draw[dashed, gray] (\xN,-2.5) -- (\xN,-84);
\draw[dashed, gray] (\xC,-2.5) -- (\xC,-84);
\draw[dashed, gray] (\xP,-2.5) -- (\xP,-84);
\draw[dashed, gray] (\xS,-2.5) -- (\xS,-84);
\draw[dashed, gray] (\xD,-2.5) -- (\xD,-46);

\draw[tcp] (\xN,-8) -- (\xC,-8)
  node[midway, above] {\textsc{Register\_Newcomer}};
\draw[tcp] (\xC,-11.5) -- (\xN,-11.5)
  node[midway, above] {snapshot};
\node[act] at (\xN,-16) {open speculative\\endpoints; wait};

\node[act, minimum width=36mm] at ({(\xP+\xD)/2},-25)
  {rescale triggered: load balancing\\evacuates departing PEs};

\draw[tcp] (\xP,-31) -- (\xC,-31)
  node[midway, above] {\textsc{Query\_Pending}};
\draw[tcp] (\xC,-34.5) -- (\xP,-34.5)
  node[midway, above] {count};
\draw[tcp] (\xP,-38.5) -- (\xC,-38.5)
  node[midway, above] {\textsc{Commit} \{kills, take\}};
\draw[tcp] (\xC,-43) -- (\xD,-43)
  node[pos=0.75, above] {\textsc{Die}};
\node[font=\tiny] at (\xD,-45.5) {exit};
\draw (\xD-1.4,-44.4) -- (\xD+1.4,-47.2);
\draw (\xD-1.4,-47.2) -- (\xD+1.4,-44.4);
\draw[tcp] (\xC,-47.5) -- (\xN,-47.5)
  node[midway, above] {\textsc{Integrate} \{id, view\}};
\draw[tcp] (\xC,-51) -- (\xP,-51)
  node[midway, above] {\textsc{Commit\_Reply}};

\draw[ucx] (\xP,-55) -- (\xS,-55)
  node[midway, above] {view delta (tree)};
\node[act] at (\xN,-60.5) {reconcile\\speculative eps};
\node[act, minimum width=28mm] at ({(\xP+\xS)/2},-60.5)
  {reconcile endpoints};

\draw[thick] (\xN-5,-66.5) -- (\xS+4,-66.5);
\node[fill=white, inner sep=1pt] at ({(\xN+\xS)/2},-66.5)
  {tree barrier over new endpoints (UCX)};

\node[act, minimum width=28mm] at ({(\xP+\xS)/2},-71.5)
  {\texttt{longjmp}; reinitialization};
\draw[ucx] (\xP,-77) -- (\xS,-77);
\draw[ucx] (\xP,-79.5) -- (\xN,-79.5)
  node[pos=0.35, above] {read-only and group state};

\draw[tcp] (\xN-4,-94) -- (\xN+6,-94) node[right] {TCP};
\draw[ucx] (\xN+16,-94) -- (\xN+26,-94) node[right] {UCX};
\end{tikzpicture}
\caption{Message sequence of a combined rescaling operation that removes the departing PEs and admits a newcomer in a single commit. Thin arrows are TCP messages to and from the coordinator; thick arrows are UCX messages. The coordinator exchanges messages only with PE~0, the departing PEs, and the newcomer; survivors receive the new view, the barrier, and the state broadcast over UCX. Newcomer registration and speculative endpoint setup overlap with normal application execution.}
\label{fig:protocol}
\end{figure}

Expansion shares the same structure with two additions (Figure~\ref{fig:protocol}). First, newcomers are spawned by the resource manager over SSH on the new hosts as soon as replacement capacity becomes available. They register with the coordinator, build speculative endpoints, and wait, all while the application continues to run. Second, when the rescaling operation is triggered, PE~0's \textsc{Query\_Pending} message returns the number of queued newcomers, and its \textsc{Commit} message absorbs the minimum of the requested and queued counts. The coordinator's \textsc{Integrate} message delivers each newcomer its node ID and the final view. Each newcomer reconciles its speculative endpoints against the final view and joins the survivors at the tree barrier.

As in the kill-and-restart model, a newcomer does not execute \texttt{main()}: it adopts the read-only data and group state that PE~0 broadcasts after its own \texttt{longjmp}. The post-expand resume callback triggers a load balancing step that populates the new PEs with migrated chares.

\subsection{Coordinator Availability and Security}
\label{sec:coord-limits}

The coordinator and PE~0 are single points of failure in the current implementation. We mitigate this risk by placing both on an on-demand instance (Section~\ref{sec:ccm-integration}), so they are exposed to ordinary hardware and software failures but not to spot interruptions, the failure mode that elasticity targets. The coordinator is also off the data path: it carries no application traffic, holds no application state, and participates only in the initial launch, rescaling commits, and the shutdown barrier. Its canonical state, an epoch and a member list, is small, and every survivor already holds a copy as its current view, so a restarted coordinator could be reconstructed from PE~0's view, with pending newcomers simply re-registering; if stronger availability is required, the view could be kept in a replicated store. Recovering from the loss of PE~0 or of the on-demand instance itself is a fault tolerance problem rather than an elasticity problem, and combining the mechanism with the existing checkpoint/restart support of Charm++ for such failures is future work.

The coordinator protocol is currently unauthenticated and unencrypted. The coordinator accepts a commit only from a connection registered as a member at the current epoch, but registration itself is not authenticated: any process that can reach the coordinator's port can register as a newcomer and be admitted into the job, with the full member list, at the next expansion. We rely on network-level isolation: in our deployment, the coordinator and all ranks run in an EC2 security group that admits traffic only from other members of the group, apart from SSH. Stronger protection, such as a per-job secret carried in every protocol message or TLS on coordinator connections, is straightforward to add; because the coordinator exchanges only small messages, over connections established outside the rescaling critical path, neither would materially affect rescaling cost.

\section{State Management Across Reinitialization}
\label{sec:state}

While the rescaling mechanism of Section~\ref{sec:design} is structurally simple, preserving the correctness of a mature, production-quality runtime system across it constitutes the principal engineering challenge of this work. When a survivor transfers control back to the runtime entry function and re-executes runtime initialization, it runs an initialization path that was written under the assumption of executing exactly once per process lifetime. Every subsystem touched by that path holds state that falls into one of two categories. Each category requires a distinct treatment, as described below.

\subsection{Two Categories of Runtime State}
\label{sec:categories}

The first category is \emph{initialization-time state} that must be preserved: handler tables, registered readonly, group, and chare metadata, CCS handler registrations, and timer epochs. Re-executing the code that initializes this state would leak, reorder, or destroy live registrations. To control this, we use a single global flag set when \texttt{setjmp} returns a nonzero value, that is, on every rescaling restart. Initialization sites that create this category of state check the flag and skip their work.

The second category is \emph{topology-dependent state} that must be adjusted after rescaling: any state that encodes the PE count, a PE numbering, a spanning tree, or the progress of an in-flight collective operation. Preserving such state unmodified would leave survivors addressing removed PEs or waiting for contributions that can never arrive. Each such subsystem exposes a rescaling reset method, invoked on every PE from a single post-restore hook. This hook executes at the point where PE~0's broadcast has restored the read-only data and group tables, so the full runtime is available but the application has not yet resumed.

The design must also maintain symmetry between survivors and newcomers. Message handlers are identified by small integer indices assigned in registration order, so survivor reinitialization must execute the same handler registration sequence as newcomer initialization. Skipping a registration on survivors but not on newcomers, or the reverse, silently misaligns every subsequent handler index between the two, and the first message dispatched across the mismatch corrupts the receiving process. This constraint determined where flag checks could be placed: registration sequences always execute, and only the mutation of already populated tables is skipped, with handler slot assignment made idempotent when a slot is already populated.

\subsection{Treatment of Runtime States}
\label{sec:cases}

Table~\ref{tab:state} summarizes the required state treatments. We describe the most instructive cases in detail.

\begin{table}[t]
\centering
\caption{Runtime state requiring explicit treatment across a rescaling operation.}
\label{tab:state}
\scriptsize
\begin{tabular}{p{0.30\columnwidth}p{0.58\columnwidth}}
\toprule
Subsystem & Treatment \\
\midrule
Handler tables, CCS table & Preserve; skip reinitialization on survivors; idempotent slot assignment \\
Reduction manager (per group) & Reset in-flight state; preserve sequence numbers; rebase contribution counts on shrink; rebuild tree with explicit activation \\
Location manager & Clear location cache; recompute home PEs; re-key local records; re-announce locations to home PEs \\
Array element tables & Re-key per-array and per-PE element hash tables under re-encoded identifiers \\
Message dispatch & Buffer arrivals from faster-restoring peers during the restore window; dispatch after restore completes \\
Load balancer & Clear in-progress flags; drain buffered rescaling requests; reset migration counters; reset synchronization barrier; propagate barrier epoch to newcomers \\
Timers & Preserve high-resolution clock epoch across rescaling operations \\
Topology caches & Invalidate caches encoding the old PE count; skip hardware topology rediscovery \\
Exit-path flags, argument vector & Reset and rebuild on each rescaling operation \\
\bottomrule
\end{tabular}
\end{table}

\subsubsection{Reduction trees} The Charm++ reduction subsystem maintains, on each PE, a local contributor count, a count of expected child contributions, and a monotonically increasing reduction sequence number per contributor. Three distinct hazards arise. First, the sequence numbers represent application progress and must survive: flushing them, as a fresh boot would, desynchronizes survivors from the reduction sequence the application is executing. Second, the expected contribution count is itself distributed state: each PE carries a share of the global contributor count, and migration deliberately leaves this share in place when a contributor moves, with the total reconciled at the root. On a shrink, the departing PEs' shares vanish with them while their former contributors, evacuated by the load balancing step, contribute from surviving PEs; each survivor must therefore rebase its share to its local contributor count, or the root observes more contributions than it expects and aborts. Third, when the reduction tree is rebuilt over the new PE numbering, a leaf with zero local contributors, typically an expansion newcomer that has not yet received migrated chares, must nevertheless actively inform its new parent that it has no outstanding contributions. Without this notification, the parent waits indefinitely and every subsequent reduction stalls.

\subsubsection{Location management} The location manager maps chare indices to home PEs and caches known locations. Both are functions of the PE count and numbering, so after a rescaling operation a survivor's tables are incorrect in every entry: cached locations refer to removed or renumbered PEs, and hashed records are keyed under stale identifier encodings. The reset clears the location cache, recomputes each local chare's home PE under the new numbering, re-keys the local records, and re-announces locations to the new home PEs. Because survivors resume independently, a location request can reach a home PE before the corresponding announcement does; the home therefore buffers requests for identifiers it does not yet know and replies once the announcement arrives, rather than dropping the request and leaving the requester's messages buffered forever. Omitting any of these steps produces sends to removed endpoints or, more subtly, messages buffered indefinitely while awaiting an element that is present but keyed differently.

\subsubsection{Message dispatch during the restore window}\label{sec:restorewindow} Survivors restore at different rates, and no barrier separates the end of one survivor's restore from another's. A fast survivor therefore begins sending post-rescale traffic, especially the location announcements and requests described above, while slower survivors are still rebuilding their state. The shutdown path routes incoming messages to a discard handler, which is correct under kill-and-restart, where a message arriving during teardown belongs to the terminated execution, but silently destroys valid post-rescale traffic under the no-restart model. The solution reuses the runtime's own boot machinery: normal startup already buffers messages that arrive before initialization completes and dispatches them once it finishes, so the re-entry point installs the same buffering handler in place of the discard handler, and the post-restore hook drains the buffer before the application resumes. The transport re-initialization performs a barrier over the new membership immediately before control returns to the runtime entry point, so every survivor has installed the buffering handler before the first post-rescale message can be transmitted.

\subsubsection{Load balancer synchronization} The load balancer's synchronization barrier and central strategy carry per-step state: an in-progress flag, migration completion counters, and a barrier epoch. The rescaling path exits in the middle of a load balancing step, and the \texttt{longjmp} bypasses the code that would normally complete the barrier, so survivors would otherwise resume with a permanently disabled barrier or with counters that either satisfy the next step's completion test spuriously or can never satisfy it. All of this state is reset in the post-restore hook. The barrier epoch must additionally be propagated to expansion newcomers: a newcomer starting at epoch zero while survivors are at epoch $k$ has its barrier contributions discarded as stale, so the cluster hangs on the second load balancing step after the expansion. Newcomers therefore receive the current epoch during integration.

\subsection{Message-Layer Identity}
\label{sec:tags}

UCX endpoints do not carry sender identity. Charm++ encodes the source node ID in the upper bits of each message tag, with the message type in the lower bits. Receive buffers are posted with a tag mask that matches only the message-type bits, so a posted receive accepts messages from any sender, and the receiver recovers the source from the matched tag. This sender-agnostic matching makes the receive path invariant under renumbering: receive buffers posted before a rescaling operation remain valid after it and are left in place. Only senders must adopt the new numbering, and in-flight traffic is drained before the reconfiguration, so no message carrying an old source ID crosses a rescaling operation; traffic sent after it that arrives while a survivor is still restoring is buffered rather than dispatched (Section~\ref{sec:restorewindow}). The compact renumbering itself is computed identically on every PE: the removal set is broadcast, and each PE derives its own new identifier, and those of all other PEs, deterministically.

\subsection{GPU State Preservation}
\label{sec:gpu-free}

Under the kill-and-restart model, GPU elasticity required a per-GPU daemon that held device buffers across the process boundary using CUDA IPC (Section~\ref{sec:killrestart}). Under the no-restart model, this subsystem is unnecessary and has been removed. A survivor's CUDA context, device allocations, streams, and registered host memory all persist unmodified across the \texttt{longjmp}; from the perspective of the device, the rescaling operation is an ordinary pause in kernel submissions. Device data moves only when the load balancer migrates a chare between PEs, in which case the existing Charm++ GPU messaging path applies, using GPUDirect RDMA where supported and host staging otherwise. Newcomer processes initialize CUDA once at spawn time, off the critical path, while awaiting integration.

Beyond the reduction in wall-clock overhead (Section~\ref{sec:evaluation}), the reduction in system complexity is substantial: there is no daemon lifecycle to manage on each host, no daemon launch on replacement instances, no IPC handle exchange, and no failure modes arising from version or timing skew between the daemon and the application.

\section{Spot Instance Integration}
\label{sec:spot}

\subsection{CharmCloudManager with No-Restart Rescaling}
\label{sec:ccm-integration}

We integrate the no-restart mechanism into \texttt{CharmCloudManager}~\cite{bhosale2025hpdc,bhosale2026ashes}. The overall structure of the framework is unchanged: an execution task creates the EC2 fleet and launches the application, and a monitoring task watches for rebalance recommendations, interruption notices, and newly launched instances. As in prior work, the fleet mixes on-demand and spot instances. We assume throughout that at least one on-demand instance is present and that PE~0 resides on it, so that the rank driving every reconfiguration, and the coordinator alongside it, are never themselves interrupted; only spot instances are candidates for removal. The rescaling signal path, however, is new. Where the restart-based integration rewrote the nodelist and invoked \texttt{charmrun} again at the new PE count, the manager now operates as follows:

\begin{enumerate}
\item It launches the coordinator alongside the job when the fleet is created.
\item When new capacity becomes available, it connects to each replacement instance over SSH and spawns the application binary with the \texttt{+newcomer} flag and the coordinator address. The newcomers register with the coordinator, initialize CUDA in the case of GPU jobs, build speculative endpoints, and wait, while the application continues to run at full capacity.
\item When it decides to rescale, it delivers a keep/remove bitmap to PE~0 over CCS. At the next load balancing step, the runtime evacuates the departing PEs and then executes the reconfiguration described in Section~\ref{sec:design}, which itself costs approximately 10\,ms.
\item It terminates vacated instances after their PEs have exited.
\end{enumerate}

Launcher independence (Section~\ref{sec:bootstrap}) is what makes the final step safe: with \texttt{charmrun\_ssh} and coordinator bootstrap, no supervised daemon runs on the vacated host, so its disappearance is not interpreted as a job failure. The initial launch, every expansion, and every shrink use the same binary and the same coordinator; no component of the job is ever relaunched.

Removals and additions can travel in the same \textsc{Commit} message, so an interruption for which replacement capacity is available is handled by a single rescaling operation combining the shrink and the expansion.

\subsection{Capacity Rebalancing with Low-Cost Rescaling}
\label{sec:rebalancing}

We retain the proactive replacement policy of prior work~\cite{bhosale2026ashes}: upon a rebalance recommendation, the EC2 signal indicating elevated interruption risk, the fleet launches replacement capacity, and the manager defers the rescaling operation until all at-risk instances have been replaced, an interruption notice forces an emergency rescale, or a timeout expires. The policy exists to consolidate replacements into few rescaling operations.

Low-cost rescaling changes this design space in two ways. First, the motivation for consolidation weakens: the consolidation window traded idle replacement instances, paid for but unused, against additional multi-second rescaling operations, and with the mechanism cost reduced to approximately 10\,ms and load balancing the only substantial per-operation cost, rescaling once per replacement as soon as it is ready becomes practical, eliminating idle time while maintaining target capacity. Second, the emergency path can act later. Under kill-and-restart the shrink itself consumed seconds of the two-minute warning window, forcing the manager to act early; with the mechanism cost negligible, the manager can wait late in the window for replacements to arrive, increasing the probability that an interruption is absorbed as a single combined operation rather than a shrink followed by a later expansion. In both directions the rescaling schedule becomes governed by capacity alone: rescale as soon as a replacement is ready, and wait when none is. We leave the exploration of both policy refinements to future work.

\section{Evaluation}
\label{sec:evaluation}

We evaluate the system along four axes: (1)~a microbenchmark decomposition of the no-restart rescaling operation; (2)~a comparison against the three interruption-handling modes measured in prior work~\cite{bhosale2026ashes}; (3)~end-to-end application runs on spot fleets with injected interruptions; and (4)~a rescaling frequency study that examines the policy space enabled by low-cost rescaling.

\textbf{Experimental setup.} We use the UCX machine layer of Charm++ with the no-restart extensions (the \texttt{ucx-linux-arm8} build on the CPU fleets and the \texttt{ucx-linux-x86\_64-cuda} build on the GPU fleets) over UCX 1.22.0 and UCX 1.20.0 respectively, both built from the UCX master branch (revisions \texttt{3de43c7} and \texttt{b9b9535}). CPU experiments use EC2 fleets of c6gn.large instances (two vCPUs each) running Jacobi2D, a communication-intensive five-point stencil benchmark; GPU experiments use g4dn.xlarge instances (one NVIDIA T4 each) running a CUDA implementation of the same computational pattern whose chare state resides in device memory. The fleets match those of prior work~\cite{bhosale2026ashes}, both benchmarks run a $16{,}384 \times 16{,}384$ grid decomposed into $2{,}048 \times 2{,}048$ blocks, and both use the GreedyRefine load balancing strategy. Rescaling requests are delivered to PE~0 over CCS by the fleet manager of Section~\ref{sec:spot}, and spot interruptions are injected with the AWS Fault Injection Simulator.

\textbf{Configuring applications for elastic execution.} No per-size configuration is required: the application is written and launched once, and the same binary, decomposition, and chare count are used at every fleet size. Two constraints bound the range of sizes a run can span. At the largest expected size, the decomposition should provide several chares per PE, so that load balancing can distribute work evenly and so that a change in PE count does not leave some PEs with an extra chare, which becomes a significant imbalance when the number of chares per PE is small. At the smallest expected size, the aggregate application state must fit in the memory of the surviving instances, including device memory for GPU applications, since evacuated chares are held in memory rather than written to storage. Within these bounds, the chare count trades load balancing granularity against per-chare overhead, as in any overdecomposed Charm++ application.

\subsection{Rescaling Overhead Breakdown}
\label{sec:eval-micro}

\begin{figure*}
\centering
\includegraphics[width=\textwidth]{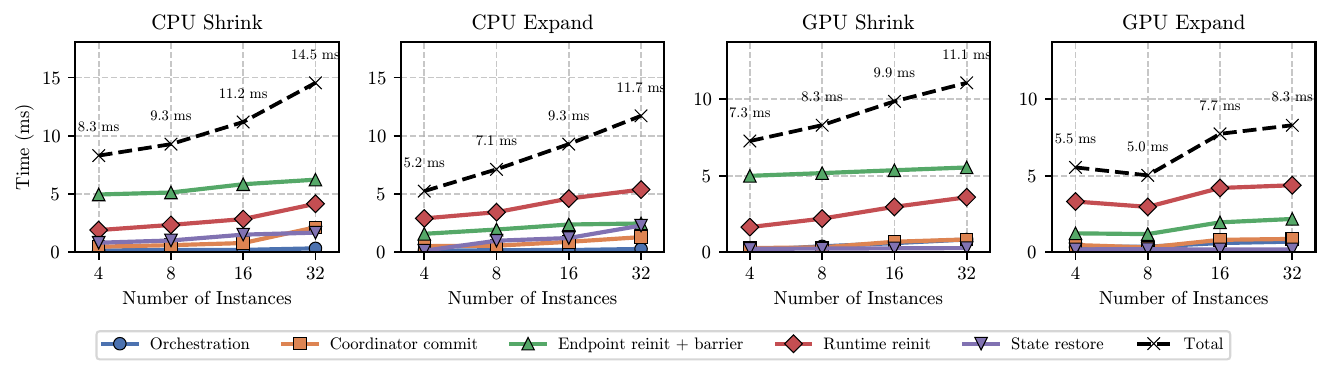}
\caption{No-restart rescaling time, excluding load balancing, versus instance count, shrinking by one instance and expanding back, on c6gn.large fleets (2 PEs per instance) and g4dn.xlarge fleets (one NVIDIA T4 per instance).}
\label{fig:breakdown-scaling}
\end{figure*}

Figure~\ref{fig:breakdown-scaling} decomposes the rescaling operation and shows how it scales on the CPU and GPU fleets, shrinking by one instance and expanding back onto it at 4 to 32 instances. The measurements exclude the load balancing step that precedes a shrink and follows an expansion, which is unchanged from the kill-and-restart model; its cost is included in the comparisons that follow. Endpoint closes are issued without blocking and complete during the post-rescale tree barrier, so the two are reported as one stage; the coordinator commit is a single round trip on PE~0, which, by the design of Section~\ref{sec:coordinator}, is independent of the number of survivors. On CPUs, total time grows from 8.3\,ms to 14.5\,ms for a shrink and from 5.2\,ms to 11.7\,ms for an expansion across an eightfold increase in fleet size: each doubling adds one to three milliseconds rather than multiplying the cost. The composition differs between the two operations. A shrink is dominated at every scale by endpoint reconciliation, which is nearly flat (5.0 to 6.2\,ms) because the number of endpoints to close depends on the membership change, not the fleet size. An expansion instead is dominated by runtime initialization (2.9 to 5.4\,ms), which the newcomer must execute in full while survivors skip it, and pays roughly 2\,ms of endpoint work since it only adds endpoints. Orchestration remains below 0.35\,ms throughout.

On GPUs, running the Jacobi2D benchmark, a shrink costs 7.3\,ms at 4 instances and 11.1\,ms at 32, an expansion 5.5\,ms and 8.3\,ms respectively, with similar stage decomposition. This is the central consequence of the no-restart design for accelerated workloads. Under kill-and-restart every process is replaced, so each rescaling operation pays CUDA context reinitialization on every surviving GPU, which cost 7 to 8\,s per operation and motivated the checkpoint daemon of prior work (Section~\ref{sec:gpu-free}). Here survivors never lose their context, and the only GPU-specific cost, context creation on a newcomer, is overlapped with the newcomer's wait for the membership commit. That overlap is essential: without it the context creation falls on the critical path of every survivor's restore, and an expansion costs 172\,ms at four instances instead of 5.5\,ms. The result is a three orders of magnitude reduction, and GPU elasticity that is no more expensive than CPU elasticity.

These measurements change membership by a single instance. For a change of $k$ instances, the coordinator round trip, the tree broadcast, and the tree barrier are unaffected, whereas endpoint reconciliation grows with $k$: each survivor closes one endpoint per departing instance and opens one per joining instance. The end-to-end runs of Section~\ref{sec:eval-endtoend} exercise such larger changes, removing and admitting up to eight instances in a single commit, and the overheads reported there include the full cost of the mechanism.

\subsection{Comparison with Restart-Based Handling}
\label{sec:eval-comparison}

We compare four modes of handling a spot interruption, the first three of which were measured in prior work~\cite{bhosale2026ashes}: (A)~checkpoint/restart through a shared filesystem (AWS EFS); (B)~in-memory kill-and-restart without capacity rebalancing; (C)~in-memory kill-and-restart with capacity rebalancing; and (D)~the mechanism presented in this paper.

\begin{figure*}
\centering
\includegraphics[width=\textwidth]{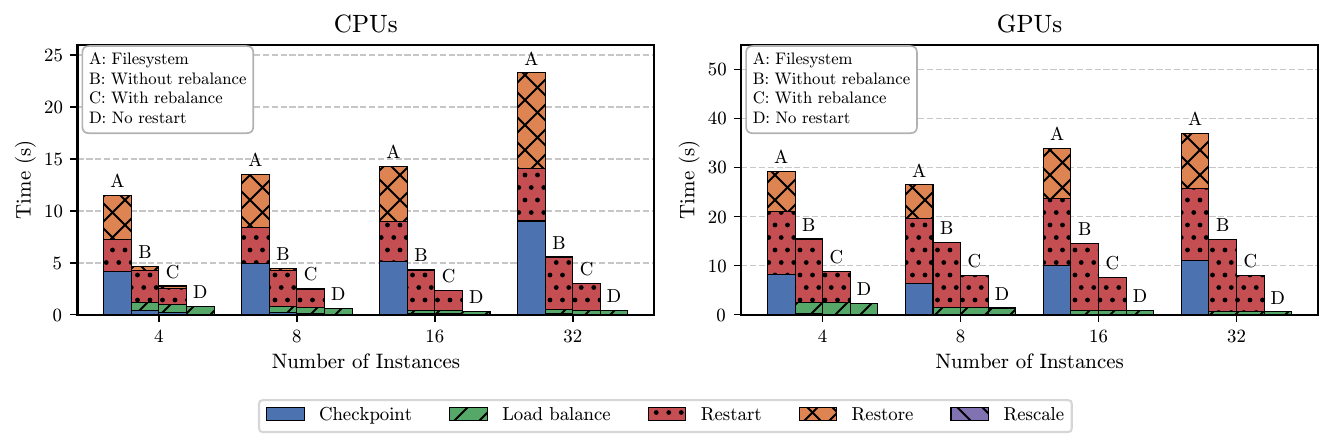}
\caption{Overhead of handling one spot interruption versus instance count, for the four handling modes, on CPU and GPU fleets. Modes A--C are the measurements of prior work~\cite{bhosale2026ashes}.}
\label{fig:mode-comparison}
\end{figure*}

Figure~\ref{fig:mode-comparison} shows the comparison. On CPUs, mode~A costs 11--23\,s depending on scale, mode~B costs approximately 4.5--5.5\,s (a shrink followed by an expansion), and mode~C costs approximately 2.2--2.9\,s (one combined rescaling operation). Mode~D performs the same single combined operation in 8--15\,ms plus the load balancing time, a reduction of more than two orders of magnitude in the mechanism cost relative to mode~C. What remains of the interruption-handling overhead in mode~D is load balancing, which is common to modes~B--D and independent of the rescaling mechanism; total handling time is $3.5\times$ lower than mode~C at four instances and $7.6\times$ lower at 32, and the rescaling component itself is no longer visible at the scale of the figure.

The GPU comparison is more favorable, because the restart term that dominates modes~A--C is larger there: CUDA context reinitialization in every replaced process costs 6.4--7.4\,s in mode~C and 12.8--14.7\,s in modes~A and~B. Mode~D removes that term entirely rather than reducing it, since survivors never leave their process and the newcomer's context creation is overlapped with its wait for the membership commit. A GPU rescaling operation costs 7--11\,ms plus the load balancing time, and total interruption handling is $3.9\times$ lower than mode~C at four instances and $13.5\times$ lower at 32, the advantage growing with scale as the load balancing term shrinks.

Two structural differences merit attention beyond the aggregate numbers. First, in modes A--C the dominant cost, restart, grows with job size, whereas in mode~D the dominant cost, endpoint reconciliation, grows only with the size of the membership change, and the remaining costs are constant. Second, mode~D eliminates the GPU checkpoint daemon entirely: the device-side checkpoint and restore stages of the prior GPU pipeline do not exist in this design, nor does the operational requirement of running daemon processes on every host.

\subsection{End-to-End Runs with Injected Interruptions}
\label{sec:eval-endtoend}

\begin{figure*}
\centering
\includegraphics[width=\textwidth]{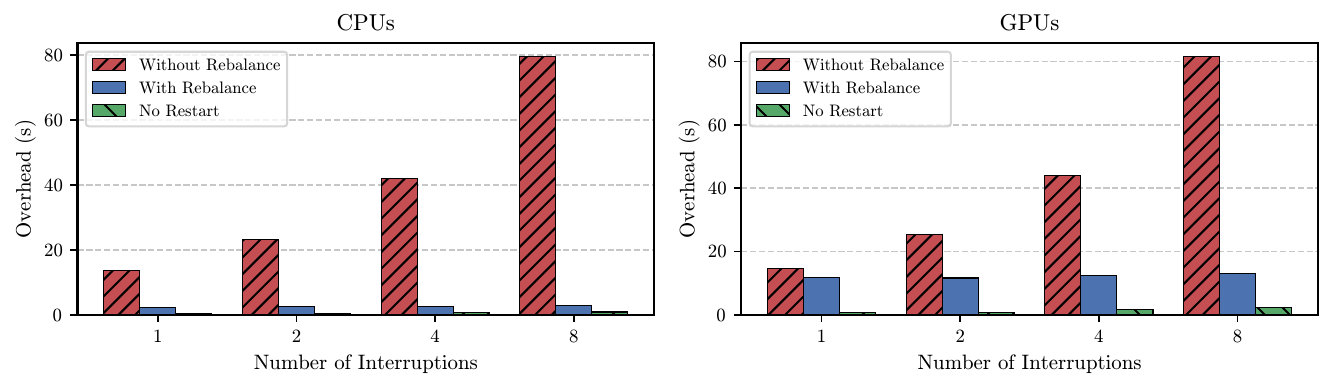}
\caption{End-to-end overhead over an uninterrupted run versus number of simultaneous spot interruptions, on 16 instances, for kill-and-restart handling without and with capacity rebalancing~\cite{bhosale2026ashes} and for the no-restart mechanism.}
\label{fig:endtoend}
\end{figure*}

Under the kill-and-restart system, eight simultaneous interruptions increased end-to-end runtime by approximately 16\% on CPUs and 20\% on GPUs without rebalancing, and by less than 1\% and 3.4\% respectively with rebalancing~\cite{bhosale2026ashes}; the residual overhead in the rebalancing case derives almost entirely from the single remaining restart. We repeat this experiment under the no-restart mechanism, with each batch of simultaneous interruptions handled by a single consolidated rescaling operation. Figure~\ref{fig:endtoend} shows the overhead each mode adds to an uninterrupted run. For eight simultaneous interruptions, the overhead falls from 2.9\,s to 0.9\,s on CPUs (0.6\% to 0.2\% of runtime) and from 13.2\,s to 2.3\,s on GPUs (3.4\% to 0.6\%). Immediate per-replacement rescaling, which trades up to $2k$ rescaling operations for $k$ interruptions against the elimination of replacement instance idle time, becomes viable at this operating point; evaluating it requires interruptions staggered in time rather than simultaneous, and we leave it to future work.

\subsection{Rescaling Frequency Study}
\label{sec:eval-frequency}

Finally, we examine how frequently a job can be rescaled before the overhead becomes significant. We run Jacobi2D on CPUs and GPUs at a fixed problem size on 16 instances and force a rescaling operation, at a fixed period, sweeping the period, and measure the aggregate slowdown relative to a run with no rescaling.

\begin{figure}
\centering
\includegraphics[width=\columnwidth]{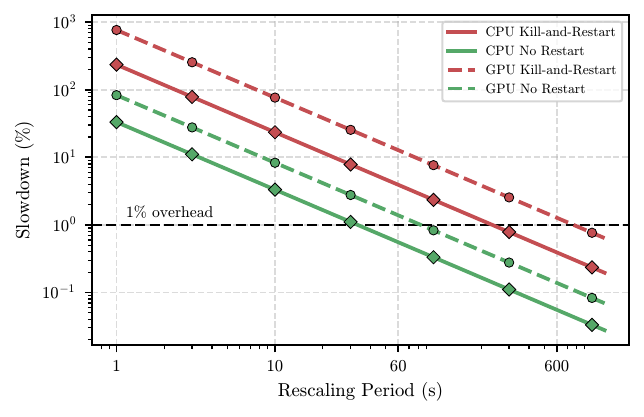}
\caption{Aggregate slowdown versus rescaling period on 16 instances.}
\label{fig:frequency}
\end{figure}

Figure~\ref{fig:frequency} shows the resulting slowdown, and the 1\% line makes the comparison concrete: to keep overhead below 1\%, kill-and-restart can sustain one rescaling operation every four minutes on CPUs and one every thirteen minutes on GPUs, whereas the no-restart mechanism sustains one every 33 seconds on CPUs and one every 83 seconds on GPUs, a $7$--$9\times$ increase in sustainable frequency.

\section{Related Work}
\label{sec:related}

Several papers have studied the performance of MPI on cloud platforms and on cloud interconnects such as AWS EFA~\cite{chakraborty2019efa,xu2020mpicloud,xu2022arm}, and the usability of cloud platforms for HPC proxy applications has been evaluated across providers~\cite{sochat2025usability}. Work targeting spot instances has largely relied on checkpoint/restart to survive interruptions~\cite{yi2010spot,yi2012spot}, including cost-aware checkpoint scheduling, or on replicated execution and hybrids of the two approaches~\cite{gong2015monetary}. Wu et al.~\cite{wu2024cantbelate} propose scheduling policies that exploit spot instances while meeting deadlines, achieving up to 84\% cost savings. Bhosale et al. ran Charm++ applications on CPU spot fleets using in-memory kill-and-restart rescaling~\cite{bhosale2025hpdc} and extended the approach to GPUs with proactive capacity rebalancing~\cite{bhosale2026ashes}. The present work removes the restart stage that dominated the remaining overhead in both.

Experimental elastic job support in Charm++ dates to Gupta et al.~\cite{gupta2014malleable}, which introduced the kill-and-restart model that this paper replaces. In the MPI ecosystem, elasticity efforts include DMR and related frameworks layered over process managers~\cite{iserte2020dmr}, the performance-driven reconfiguration of Flex-MPI~\cite{martin2015flexmpi}, and API extensions that grow and shrink the world communicator within adaptation windows~\cite{compres2016elastic}. MPI Sessions~\cite{holmes2016sessions} relaxes the static-world assumption of the MPI standard, and recent work builds dynamic resource management on Sessions and PMIx~\cite{huber2022sessions}. However, production MPI reconfiguration continues to require either a job restart or the cooperation of a supervised process manager, which is precisely the component that spot interruptions disrupt. Elastic execution has also been studied in Kubernetes-based schedulers for HPC and machine learning workloads~\cite{milroy2022converged,misale2021kubeflux,hsieh2023voda,bhosale2025elastic}, which rescale at the granularity of pods and rely on application-level checkpoint/restart. Our mechanism is complementary to these systems: it provides an in-application rescaling path that such orchestrators could drive in place of pod-level checkpoint/restart.

The closest technical relatives of our reinitialization approach come from MPI fault tolerance. Fenix~\cite{gamell2014fenix}, built on the failure-repair primitives of ULFM~\cite{bland2013ulfm}, and Reinit++~\cite{georgakoudis2020reinit} implement global-restart recovery in which surviving processes are not terminated: after a failure, execution returns to an initialization point in each live process, the MPI runtime is repaired or reinitialized, and application state is rolled back from checkpoints. These systems establish the value of avoiding job redeployment, but they recover a job of unchanged size, roll back state rather than preserving it, and operate on the process manager's supervised infrastructure. The no-restart mechanism extends the surviving-process model from failure recovery to deliberate reconfiguration: membership, numbering, and topology-dependent state change; live state is preserved without rollback; and no supervised infrastructure is required. To our knowledge, no prior tightly coupled HPC runtime system performs a rescaling operation without restarting the surviving processes.

\section{Conclusion}
\label{sec:conclusion}

This paper presented a no-restart rescaling mechanism for the Charm++ runtime system, in which surviving processes reconfigure in place rather than terminating and restarting at the new processor count. Excluding the load balancing step common to both models, a rescaling operation costs 8 to 15\,ms on CPU fleets and 7 to 11\,ms on GPU fleets between 4 and 32 instances, two to three orders of magnitude below the multi-second cost of restart-based implementations, and grows by one to three milliseconds per doubling of the fleet. Because processes survive, device state persists in place and the checkpointing daemons previously required for GPU elasticity are eliminated. Integrated with a spot instance management framework, the mechanism reduces the end-to-end overhead of eight simultaneous interruptions from 0.6\% to 0.2\% of runtime on CPUs and from 3.4\% to 0.6\% on GPUs, and increases the rate at which a job can be rescaled below 1\% overhead by a factor of seven on CPUs and nine on GPUs.

Although the implementation targets Charm++, only part of it is specific to Charm++. The coordinator and the SSH launcher have no Charm++ dependencies, and endpoint reconciliation depends only on the UCX transport. What the mechanism requires of the runtime system is migratable application state, so that departing processes can be evacuated, and an initialization path that can be re-entered. Other runtime systems with these properties, such as asynchronous many-task runtimes or MPI implementations with global-restart recovery~\cite{georgakoudis2020reinit}, could adopt the same approach.

\section*{Acknowledgments}

Claude Fable was used to generate and to vet the logic of Figures~\ref{fig:components} and~\ref{fig:protocol}, which illustrate the system components and the rescaling protocol of Section~\ref{sec:design}.

\bibliographystyle{IEEEtran}
\bibliography{references}

\clearpage

\appendix[Artifact Description]

\section*{Part 1: Overview of Contributions and Artifacts}

\subsection*{A. Paper's Main Contributions}

\begin{description}
  \item[$C_1$] A no-restart shrink/expand mechanism for a message-driven
    parallel runtime system, in which surviving processes reconfigure in
    place through coordinated membership change, in-place endpoint
    reconciliation, and runtime reinitialization without process exit.
  \item[$C_2$] A treatment of runtime state across reinitialization that
    separates state which must be preserved from state which must be
    adjusted, including the message buffering required while survivors
    restore at different rates.
  \item[$C_3$] A launcher-independent bootstrap built on a standalone
    membership coordinator and an SSH launcher, removing the dependence
    on supervised process managers that spot interruptions disrupt.
  \item[$C_4$] GPU elasticity without checkpointing daemons: because
    processes survive, CUDA contexts and device-resident data persist in
    place.
  \item[$C_5$] Integration with a spot instance management framework and
    an analysis of how low-cost rescaling changes viable
    interruption-handling policy.
\end{description}

\subsection*{B. Computational Artifacts}

\begin{table}[h]
\centering
\begin{tabular}{lll}
\hline
Artifact & Contributions & Paper elements \\
\hline
$A_1$: Rescaling breakdown & $C_1$, $C_2$, $C_4$ & Figures 3, 4 \\
$A_2$: Interruption runs   & $C_1$, $C_4$, $C_5$ & Figure 5 \\
$A_3$: Frequency study     & $C_1$, $C_5$        & Figure 6 \\
\hline
\end{tabular}
\end{table}

All artifacts run on the extended Charm++ runtime driven by the
\texttt{charm-aws} fleet management and experiment harness (branch
\texttt{canopie26} of both). The \texttt{charm-aws} \texttt{README}
documents the machine images and build configurations, the AWS resources
and quotas assumed, the exact command line for every experiment, and the
figure generation steps:

{\footnotesize
\begin{itemize}
  \item \url{https://github.com/charmplusplus/charm-aws/tree/canopie26}
  \item \url{https://github.com/charmplusplus/charm/tree/canopie26}
\end{itemize}
}

Every driver accepts \texttt{-{}-gpu}, which selects the GPU machine
image, instance type, benchmark, and Charm++ build, so the CPU and GPU
variants of each experiment are the same code path. Every driver also
accepts \texttt{-{}-parse-only}, which re-derives the reported numbers
from captured application output without allocating cloud resources, so
the analysis and figures can be reproduced from an archived results set
alone.

\section*{Part 2: Artifact Identification}

\subsection*{Artifact $A_1$: Rescaling breakdown}

\subsubsection*{A. Relation to Contributions}
Measures the cost of a single rescaling operation and its per-stage
decomposition against fleet size ($C_1$, $C_2$), on both CPU and GPU
fleets ($C_4$), producing Figure 3. The same measurements supply the
no-restart bars of the mode comparison in Figure 4, whose
kill-and-restart bars are the published measurements of prior work.

\subsubsection*{B. Expected Results}
Excluding load balancing, a shrink costs 8--15\,ms and an expansion 5--12\,ms on CPU fleets of 4 to
32 instances, with GPU fleets within a few milliseconds of the CPU case;
shrink is dominated by endpoint reconciliation, which is nearly flat in
fleet size, and expansion by runtime initialization on the newcomer. The
first rescaling operation of a run is several times more expensive than
the rest, so the minimum over trials is reported.

\subsubsection*{C. Expected Reproduction Time (in minutes)}
Setup 60 per platform (image preparation); execution 60 (CPU) and 60
(GPU) for the full 4/8/16/32 sweep at three trials; analysis 5.

\subsubsection*{D. Artifact Setup}\mbox{}

\emph{Hardware:} EC2 fleets of up to 32 \texttt{c6gn.large} instances
(CPU) and up to 32 \texttt{g4dn.xlarge} instances (GPU), launched into a
single availability zone with a cluster placement group.

\emph{Software:} Charm++ with the no-restart extensions, built as
\texttt{ucx-linux-arm8-openpmix} (CPU) and
\texttt{ucx-linux-x86\_64-cuda-openpmix} (GPU); UCX 1.22.0 (CPU, master revision \texttt{3de43c7}) and 1.20.0 (GPU, master revision \texttt{b9b9535}); OpenPMIx; Python~3
with \texttt{boto3}, \texttt{asyncssh}, and \texttt{matplotlib}.

\emph{Datasets/Input:} none; Jacobi2D (CPU) and Stencil2D (GPU) run a
$16{,}384 \times 16{,}384$ grid in $2{,}048 \times 2{,}048$ blocks.

\emph{Installation \& Deployment:} see the \texttt{charm-aws}
\texttt{README}. The benchmark must be relinked whenever the runtime is
rebuilt, since the source and build trees hold independent binaries.

\subsubsection*{E. Artifact Evaluation}
$T_1$ (\texttt{run\_breakdown.py} launches a fleet per instance count,
runs the application, and issues shrink and expand requests over CCS for
each trial) $\rightarrow$ $T_2$ (the runtime emits a per-stage timing
block on PE~0 for each operation, captured in the run log) $\rightarrow$
$T_3$ (\texttt{plot\_breakdown.py} and
\texttt{plot\_mode\_comparison\_norestart.py} render Figures 3 and 4).
The driver aborts if EC2 fulfills fewer instances than requested, rather
than recording a smaller cluster under the requested label.

\subsubsection*{F. Artifact Analysis}
The driver parses the timing blocks into CSV and JSON, classifying each
operation as a shrink or an expansion from the preceding coordinator
commit, and reports the fastest trial per point.

\subsection*{Artifact $A_2$: Interruption runs}

\subsubsection*{A. Relation to Contributions}
Runs the application to completion under simultaneous spot interruptions
($C_1$, $C_5$) on both platforms ($C_4$), producing Figure 5. Each batch
of interruptions is handled by one combined rescaling operation, the
capacity-rebalanced path of Section~\ref{sec:spot}.

\subsubsection*{B. Expected Results}
End-to-end overhead grows slowly with the number of simultaneous
interruptions and remains well below one percent of runtime on CPUs and
approximately one percent on GPUs at eight interruptions, with the
residual overhead attributable to load balancing rather than to the
rescaling mechanism.

\subsubsection*{C. Expected Reproduction Time (in minutes)}
Setup as $A_1$; execution 150 per platform for interruption counts
0/1/2/4/8; analysis 5.

\subsubsection*{D. Artifact Setup}\mbox{}

\emph{Hardware:} 16 instances for the job plus idle replacement capacity,
per platform, as in $A_1$.

\emph{Software, Datasets/Input, Installation:} as $A_1$.

\subsubsection*{E. Artifact Evaluation}
$T_1$ (\texttt{run\_endtoend.py} launches the fleet and starts the
application) $\rightarrow$ $T_2$ (at a fixed fraction of the run, $k$
newcomers are spawned on the replacement instances and a single client
request removes the $k$ victims and admits the $k$ replacements)
$\rightarrow$ $T_3$ (the run completes and its total execution time is
compared against the uninterrupted baseline).

\subsubsection*{F. Artifact Analysis}
The driver extracts total execution time from the application's
completion output and reports overhead against the $k=0$ run.

\subsection*{Artifact $A_3$: Frequency study}

\subsubsection*{A. Relation to Contributions}
Establishes how often a job can be rescaled before the overhead becomes
significant ($C_1$, $C_5$), producing Figure 6.

\subsubsection*{B. Expected Results}
Slowdown is inversely proportional to the rescaling period. Overhead
stays below one percent for periods of tens of seconds, roughly an order
of magnitude shorter than the periods at which restart-based rescaling
reaches the same threshold, and the residual per-operation cost is
dominated by load balancing rather than by the mechanism.

\subsubsection*{C. Expected Reproduction Time (in minutes)}
Setup as $A_1$; execution 120 per platform for the period sweep;
analysis 5.

\subsubsection*{D. Artifact Setup}\mbox{}

\emph{Hardware:} 16 instances per platform, as in $A_1$.

\emph{Software, Datasets/Input, Installation:} as $A_1$.

\subsubsection*{E. Artifact Evaluation}
$T_1$ (\texttt{run\_frequency.py} launches the fleet and starts the
application) $\rightarrow$ $T_2$ (a background task alternates shrink and
expand requests at the configured period until the run completes; a
newcomer is spawned before each expansion) $\rightarrow$ $T_3$ (execution
time and the number of committed operations are compared against a
baseline run with no rescaling).

\subsubsection*{F. Artifact Analysis}
The driver computes slowdown relative to the baseline and divides the
additional time by the number of committed rescaling operations to
recover the per-operation cost, rendered by
\texttt{plot\_frequency.py}.

\end{document}